\documentclass[12pt]{article}
\usepackage{amsmath,cite,epsf}

\newcommand{\tr}{\mathop{\rm Tr}}
\newcommand{\str}{\mathop{\rm Str}} 
\newcommand{\half}{{\textstyle\frac12}}
\newcommand{\third}{{\textstyle\frac13}}
\newcommand{\quart}{{\textstyle\frac14}}
\makeatletter
\@addtoreset{equation}{section}
\makeatother

\newcommand{\jref}[4]{{\it #1} {\bf #2}, #3 (#4)}

\newcommand{\IJMP}[3]{\jref{Int.\ J.\ Mod.\ Phys.}{#1}{#2}{#3}}

\newcommand{\PA}[3]{\jref{Physica}{#1A}{#2}{#3}}
\newcommand{\PLB}[3]{\jref{Phys.\ Lett.}{#1B}{#2}{#3}}

\newcommand{\PRD}[3]{\jref{Phys.\ Rev.}{D#1}{#2}{#3}}

\makeatletter
\def\vereq#1#2{\lower3pt\vbox{\baselineskip1.5pt \lineskip1.5pt
\ialign{$\m@th#1\hfill##\hfil$\crcr#2\crcr\sim\crcr}}}
\makeatother

\begin{document}

\begin{titlepage}
\begin{center}
{\hbox to\hsize{hep-th/9807098 \hfill   UCLA/98/TEP/32}}
{\hbox to\hsize{               \hfill  MIT-CTP-2756}}
{\hbox to\hsize{               \hfill UCSD-PTH-98-24}}

\vskip 1in
 
{\Large \bf Field Theory Tests for Correlators
\\ \vskip 0.15in
in the AdS/CFT Correspondence }

\vskip 0.45in

{\bf Eric D'Hoker${}^1$, Daniel Z. Freedman${}^2$, and Witold Skiba${}^3$}

\vskip 0.15in
{\em ${}^1$Department of Physics,
University of California, Los Angeles, CA 90024 \\ }
\vskip 0.1in
{\em ${}^{1,2}$Institute for Theoretical Physics, \\
University of California, Santa Barbara, CA 93106 \\ }
\vskip 0.1in
{\em ${}^2$Department of Mathematics and Center for Theoretical Physics,\\
Massachusetts Institute of Technology, Cambridge, MA 02139 \\ }
\vskip 0.1in
{\em ${}^3$Department of Physics, University of California at San Diego,
La Jolla, CA 92093 \\}

\vskip 0.1in

\end{center}

\vskip .5in

\begin{abstract}
The order $g^2$ radiative corrections to all 2- and 3-point correlators
of the composite primary operators $\tr X^k$ are computed in ${\cal N}=4$
supersymmetric Yang-Mills theory with gauge group $SU(N)$. Corrections
are found to vanish for all $N$\@. For $k=2$ this is a consequence of known
superconformal nonrenormalization theorems, and for general $k$ the
result confirms an $N \rightarrow \infty$, fixed large $g^2N$ supergravity
calculation and further conjectures in hep-th/9806074. 
A 3-point correlator involving ${\cal N}=4$ descendents of $\tr X^2$ is
calculated, and its order $g^2$ contribution also vanishes, giving evidence for
the absence of radiative corrections in correlators of descendent
operators.

\end{abstract}

\end{titlepage}

\newpage

\section{Introduction}
\setcounter{equation}{0}

The Maldacena conjecture~\cite{Maldacena,Polyakov,Witten}
has brought fruitful insights
into string theory, supergravity (SG), and supersymmetric field theory while
emphasizing a new and striking role of anti-de-Sitter space (AdS) in theoretical
physics. Many correlation functions [2-14] have now been computed and compared
in the boundary ${\cal N}=4$ supersymmetric Yang-Mills theory (SYM) and bulk
$AdS_5$ SG theories. Certain 2- and 3-point correlators of operators in the same
multiplet as the stress tensor and $SU(4)$ flavor currents satisfy superconformal
nonrenormalization theorems~\cite{AFGJ,GubKle} and thus have no radiative
corrections. 

In a very recent
paper~\cite{LMRS} the correlators of the chiral primary family\footnote{
$\tr X^k$ is shorthand for $\tr X^{\{ i_1} X^{i_2}\cdots X^{i_k \} }$,
an operator in the rank $k$ traceless symmetric tensor product of the
fundamental real scalar fields $X^i$ of the ${\cal N}=4$ SYM theory.
These fields are $SO(6)$ vectors in the adjoint representation of
the $SU(N)$ gauge group. It is convenient to introduce $X^i = T^a X^i_a$,
where $T^a$ is a hermitian generator of the $SU(N)$ fundamental,
with normalization $\tr T^a T^b = \delta^{ab}/2$. The structure constants
$f^{abc}$ of $SU(N)$ are normalized by $[T^a, T^b] = i f^{abc} T^c$.}
$\tr X^k$ were studied. Three-point functions
$\langle \tr X^{k_1} \, \tr X^{k_2} \, \tr X^{k_3} \rangle$
of normalized operators were  evaluated using an $N\rightarrow \infty$
fixed large $g^2 N$ supergravity computation
and shown to be equal to their free-field values. A conjecture was
made that the correlators are independent of $\lambda=g^2 N$ to leading order
in $N$ and also a final speculation that they are independent of $g^2$ even for
finite $N$. Known nonrenormalization theorems apply only to the case $k_1=k_2
=k_3=2.$  Thus, these results and speculations hint at nonrenormalization 
properties which go beyond the known symmetries of ${\cal N}=4$ SYM theory.

In this paper, we check the nonrenormalization properties
of the 2- and 3-point functions of general chiral primary operators
$\tr X^k$ to order $g^2$, by an explicit computation in the (boundary) ${\cal
N}=4$ SYM theory. A further calculation of a 3-point function involving
${\cal N}=4$ descendents of the operator $\tr X^2$ to the same order gives
evidence that nonrenormalization also holds for 
descendents of the chiral primary operators.

The essential steps in the method are:
\begin{itemize}
\item[{\it i.}]
We use an ${\cal N}=1$ description of the 6 real fields $X^i$ as 3 complex 
$z^i$ and their conjugates $\bar{z}^i$ in ${\bf 3}$ and ${\bf \bar{3}}$
representations of the manifest $SU(3)$ subgroup of the original $SU(4)$
flavor group.  Without loss of generality, a choice of highest weight operators
in the rank $k$ symmetric tensor representations of $SU(3)$ is made,
which simplifies the flavor combinatorics.
\item[{\it ii.}]
The essential part of the problem is then reduced to that of color combinatorics
which we handle by summing over all permutations of color generators in traces.
Within each such permutation of $z^i$-lines we sum over choices of pairs of lines
carrying interaction vertices and choices of lines carrying self-energy
insertions.
\item[{\it iii.}]
In Section~\ref{sec:kk}, we apply these steps to $\langle \tr X^k\,
\tr X^k\rangle$ and show that the
sum of all interactions gives an order $g^2$ amplitude proportional to that 
for the case $k=2$ which vanishes by a known nonrenormalization theorem.
\item[{\it iv.}]
In Section~\ref{sec:kk2}, we study the combinatorics of the 3-point function 
$\langle \tr X^k\, \tr X^k\, \tr X^2 \rangle$.
The net sum of interactions produces amplitudes which
again vanish by nonrenormalization theorems for $k=2$.
\item[{\it v.}]
The analysis of the general 3-point correlator
$\langle \tr X^{k_1}\, \tr X^{k_2}\, \tr X^{k_3} \rangle$ 
is somewhat more involved and is presented in Sec~\ref{sec:ijk}.
\end{itemize}

Specifically, our results are as follows. The free field contributions to the
above correlators are polynomials in
$N$ of leading order $N^k$ for 2-point functions and order
$N^{(k_1+k_2+k_3)/2-1}$ for 3-point functions. The cancellations of radiative
corrections that we find to order $g^2$ occur for all $N$, not just leading
order.  Therefore, we confirm the results and speculations of~\cite{LMRS} in
their strongest form. Concretely, the origin of the cancellations to order
$g^2$ may be traced back to the fact that  only interactions which involve
at most two $X$-lines 
appear, so that nonrenormalization theorems for
$k=2$ can be used in the proof for general values of $k$. To order
$g^4$ however, there are also interactions involving three $X$-lines,
and the situation is far more complicated. There may exist special symmetries or
properties of the ${\cal N}=4$ theory which lead to
nonrenormalization theorems in higher order, but the only present evidence is 
the large $N$ calculations of~\cite{LMRS}.

Another question of interest within the ${\cal N}=4$ SYM theory is whether the
absence of radiative corrections for 2- and 3-point functions of $\tr X^k$ 
established by nonrenormalization theorems for $k=2$ or by calculation
for general $k$ remains true for all supersymmetric partners (descendents) of
these operators. The case $k=2$ is of special interest because the multiplet of
descendents  includes the $SU(4)$ flavor currents and the stress tensor. 
In this case the formalism of ~\cite{HoweWest} predicts that there is a
unique ${\cal N}=4$ superconformal correlator for this multiplet, so
that radiative corrections of descendents should vanish as well. To check this we
note that the second descendent of $\tr X^2$ has the schematic form 
$S^{ij} = \tr \lambda^i \lambda^j + g (\tr X^3)^{ij}$, where
$\tr X^2$ transforms in the
${\bf20'}$ of $SU(4)$ while $S^{ij}$ transforms in the
${\bf 10}$. We discuss the specific form of
$S^{ij}$ in Section~\ref{sec:descendent} and show that both
free-field and order $g^2$ contributions to the correlator
$\langle S^{ij} S^{kl} \tr X^2 \rangle$ vanish.

Although our method requires knowledge of the structure of the ${\cal N}=4$
theory only at a schematic level, we record here for reference the
(Euclidean signature) Lagrangian in an ${\cal N}=1$ component notation
(where $L$ and $R$ are chirality projectors):
\begin{multline}
\label{eq:n4lag}
{\cal L}  =  \left[ \quart F_{\mu \nu }^{~~2}+\half\bar{\lambda}%
             {D}\!\!\!\!\slash\lambda
             +\overline{D_{\mu }z^i }D_{\mu }z^i +\half 
             \bar{\psi}^i{D}\!\!\!\!\slash\psi^i \right.  \\
 + i\sqrt{2} g f^{abc}(\bar{\lambda}_{a}\bar{z}_b^i L\psi_c^i
  -\bar{\psi}_a^i R z_b^i\lambda_c)   
 - \frac{Y}{2} f^{abc} (\epsilon_{ijk} \bar{\psi}_a^i L z_b^j \psi_c^k
  +\epsilon_{ijk} \bar{\psi}_a^i R \bar{z}_b^j \psi_c^k)    \\ 
 \left. -\half g^{2}(f^{abc} \bar{z}_b^i z_c^i)^{2}
  +\frac{Y^2}{4} f^{abc} f^{ade} \epsilon_{ijk} \epsilon_{ilm}
   z_b^j z_c^k \bar{z}_d^l\bar{z}_e^m\right]  
\end{multline}
We use separate coupling constants
$g$ and $Y$ in order to distinguish interactions arising from the gauge and
superpotential sectors. The relation $Y = \sqrt{2} g$ produces ${\cal N}=4$
SUSY\@. 
\section{The Correlators \protect\boldmath$\langle \tr X^k\, \tr X^k
\rangle$\label{sec:kk}}
\setcounter{equation}{0}

\begin{figure}[htb]
\centerline{\epsfysize=2.5cm\epsffile{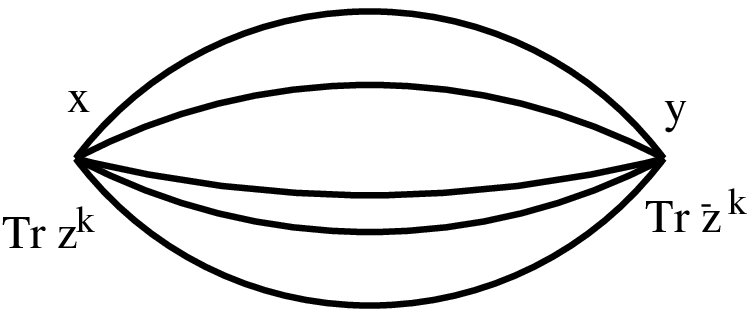}}
\caption{Born graph for the $\langle \tr(z^1)^k\, \tr (\bar{z}^1)^k \rangle$
         correlator.
         \label{fig:2Born}}
\end{figure}

Without loss of generality, we may choose to evaluate the correlators of
the $SU(3)$ flavor highest weight field $z^1$ and its complex conjugate
$\bar{z}^1$ only. This choice vastly simplifies the problem since flavor
combinatorics becomes trivial and separates from color combinatorics. We
therefore study the correlator
$\langle\tr(z^1)^k\, \tr (\bar{z}^1)^k\rangle$ for which contributing Feynman
diagrams are shown in Fig.~\ref{fig:2Born} and 2. The free field contribution 
is a sum
over permutations of ordering in the second operator trace relative to a fixed
ordering in the first operator trace,
\begin{align}
\label{eq:kktree}
\langle \tr(z^1)^k(x) \, \tr(\bar{z}^1)^k(y) \rangle
 &=  G(x,y)^k P_{k,k,0}(N)
\nonumber \\
P_{k,k,0}(N)
 &\equiv 
\sum_{{\rm perms}\,\, \sigma}\!\!
      \tr\left( T^{a_1} \cdots T^{a_k} \right) \,
      \tr\left( T^{a_{\sigma(1)}} \cdots T^{a_{\sigma(k)}} \right)
\end{align}
where $G(x,y)=1/[4 \pi^2 (x-y)^2]$ is the free scalar propagator, and the sum
in the second line is over all permutations $\sigma$. The permutation sum of
contracted traces defines the polynomial in $N$ we have called $P_{k,k,0}(N)$. By
cyclicity of the trace there are $(k-1)!$ classes of permutations each with $k$
identical contributions. So
$P_{k,k,0}(N)$ has an overall factor of k. The leading term comes from ``reverse
order permutations'' ($\sigma(1)=k$, $\sigma(2)=k-1$, etc) and is of order
$P_{k,k,0}(N) \sim k (N/2)^k$, in agreement with~\cite{LMRS}.

From the Lagrangian one can easily see that there are two-particle interactions
 inside the
``rainbow'' coming from gauge boson exchange and D-term quartic vertices. 
(F-term
vertices do not contribute.) The sum of these is a basic two particle interaction
of the following structure
\begin{align}
\label{eq:B}
    \mbox{\raisebox{-.45truecm}{\epsfysize=1cm\epsffile{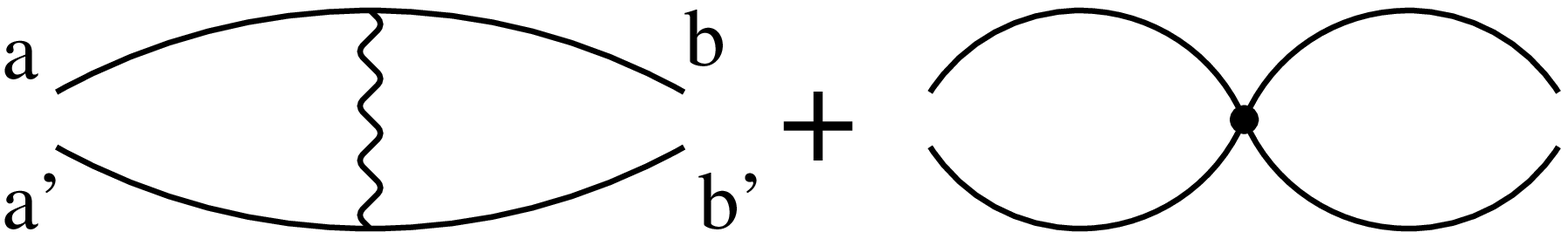} }}
 &=(f^{pab} f^{p a' b'} + f^{pab'} f^{p a' b}) B(x,y) G(x,y)^2  \\
 B(x,y) &= b_0 + b_1 \ln(x-y)^2 \mu^2 \nonumber
\end{align}
where $a, a', b, b'$ indicate adjoint color indices. The indicated space-time
dependence is the result of integrations which may be performed explicitly, but
it turns out that we will not need the precise values of $b_0,b_1$. Self-energy
corrections to order $g^2$ arise from two sources : a gauge boson exchange and a
fermion loop.  They are represented in the figures below by a dark dot, and may
be expressed as follows 
\begin{align}
\label{eq:A}
    \mbox{\raisebox{-.2truecm}{\epsfysize=0.5cm\epsffile{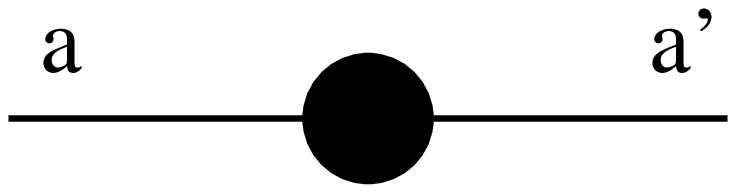} }}
    &=  \delta^{aa'} N A(x,y) G(x,y) \\
  A(x,y) &=  a_0 + a_1 \ln(x-y)^2 \mu^2. \nonumber
\end{align}
The coefficients $a_i,b_i$ are gauge
dependent, but we know that the final color traced amplitude is gauge
independent and that logarithms will also cancel, since the primary operators
cannot have anomalous dimension.
\begin{figure}[htb]
\centerline{ \epsfysize=2.5cm\epsffile{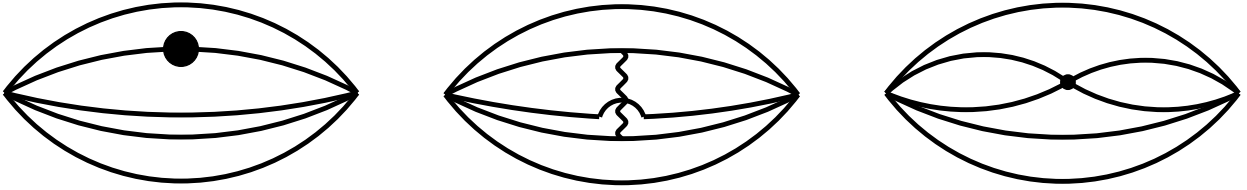}}
\caption{Order $g^2$ corrections to the 2-point function.
         \label{fig:2pointg2}}
\end{figure}

We shall now show that the sum of the $g^2$ contributions to the 2-point
functions of Fig.2 precisely cancel for all $N$ and all $k$.
The study of the color factors is facilitated by the following identity
for any set of matrices $M_i$ and $N$:
\begin{equation}
\label{eq:identity}
 \sum_{i=1}^n \tr  (M_1 \cdots [M_i,N] \cdots M_n) = 0.
\end{equation}
We will use this identity here and in following sections.

For each fixed permutation of final relative to initial $z$-lines we
insert the two particle interaction structure Eq.~\ref{eq:B} between all
distinct  pairs of lines. We then convert the two $f$ symbol
contractions to commutators within the final trace, and obtain the expression
(omitting an overall factor of $G(x,y)^k$ throughout)
\begin{equation}
\label{eq:kkfour}
\quart (-2 B) \, \tr (T^{a_1} \cdots T^{a_k})\,
\sum_{i \neq j =1}^{k} \tr  \left( T^{a_{\sigma(1)}} \cdots 
[T^{a_{\sigma(i)}}, T^p] \cdots [T^{a_{\sigma(j)}}, T^p] 
\cdots T^{a_{\sigma(k)}} 
\right).
\end{equation}
The factor of 1/4 is inserted because the expression
would otherwise overcount pairs for two reasons: first by the
expected factor of two for $\sum_{i \neq j}$ rather than $\sum_{i<j}$,
and second since a given choice of the pair $\sigma(i),\sigma(j)$ occurs
both in the fixed final trace permutation chosen and in the permutation
with $\sigma(i)$ and $\sigma(j)$ interchanged. The $-$ sign appears because 
of the $i$'s in $[T,T]=ifT$ and the factor of 2 because the two $ff$ terms in
Eq.~\ref{eq:B} contribute equally. 

We now use Eq.~\ref{eq:identity} on one of the commutators to rewrite
Eq.~\ref{eq:kkfour} as
\begin{gather}
\label{eq:kkfive}
 \half \, B \tr ( T^{a_{1}}\cdots T^{a_{k}})
\sum_{i=1}^{k} \, \tr  \left(  T^{a_{\sigma(1)}}\cdots
 [\, [ T^{a_{\sigma(i)}}, T^p ], T^p]\cdots T^{a_{\sigma(k)}} \right)
\nonumber \\
 = \half \, N B \tr ( T^{a_{1}}\cdots T^{a_{k}})
\sum_{i=1}^{k} \, \left( \tr   T^{a_{\sigma(1)}}
\cdots T^{a_{\sigma(i)}}\cdots T^{a_{\sigma(k)}} \right)
\end{gather}
where the last equality follows since $[[\ ,T^p],T^p]$ is the Casimir 
operator in the adjoint representation, so that
for any generator $T^a$ we have $[[T^a,T^p],T^p]=N T^a$. The sum $\sum_{i}$
is now over $k$ identical terms and thus collapses to an overall factor $k$.
The self-energy corrections also produce an obvious factor of $k$ relative to the
free field form. Adding interaction and self-energy contributions, we find the
following result, including the sum over all permutations~$\sigma$
\begin{equation}
\label{eq:kksix} 
\half  k N (B+2A)   \sum _{{\rm perms} \ \sigma}\!\! \tr                 
\left( T^{a_1} \cdots T^{a_k} \right) \,
   \tr    \left( T^{a_{\sigma(1)}} \cdots T^{a_{\sigma(k)}} \right).
\end{equation}
The nonrenormalization
theorem for $k=2$ implies that $B+2A=0$, and this is enough to show
that radiative corrections vanish for all values of $k$ ! Note that this
argument does not require the explicit expressions for $A$ and $B$.

\section{The Correlators \protect\boldmath$\langle \tr X^k \, \tr X^k \, \tr X^2
\rangle
$\label{sec:kk2}} 
\setcounter{equation}{0}

As we noted in the previous section, flavor combinatorics is greatly
simplified by considering the $SU(3)$ flavor highest weight fields
for the $\tr X^k$ operators and a flavor nonsinglet for the operator
$\tr X^2$. Without loss of generality, we therefore study the correlators
$\langle \tr (z^1)^k \, \tr (\bar{z}^1)^k \, \tr (\bar{z} t z) \rangle $, where
$t$ is a diagonal $SU(3)$ generator. This choice ensures that $\tr (\bar{z} t
z)$ transforms as ${\bf 8}$ of $SU(3)$, which is a part of the ${\bf 20'}$ of
$SU(4)$ and does not mix with the singlet. The free field contribution to this
correlator is a sum of all possible pairings of fields into free propagators
\begin{align}
\label{eq:kk2tree}
\langle \tr (z^1)^k(x) \, \tr (\bar{z}^1)^k(y) \, \tr
 (\bar{z} t z)(w) \rangle &=  
 G(x,y)^{k-1} G(x,w) G(y,w) P_{k,k,2}(N) t_{11} \nonumber \\
P_{k,k,2}(N)
&\equiv   k\!\! \sum_{{\rm perms}\ \sigma }\!\!
      \tr\left( T^{a_1} \cdots T^{a_k} \right) \,
      \tr\left( T^{a_{\sigma(1)}} \cdots T^{a_{\sigma(k)}} \right).
\end{align}
The additional factor of $k$ enters because of the distinguished line to
the third  vertex $\tr (\bar{z} t z)$. This line can appear at any
position in the first trace, and may then be moved to first position using
cyclicity.
Since $P_{k,k,2} = k P_{k,k,0}(N)$, the leading term for large $N$ is
$P_{k,k,2}\sim k^2 (N/2)^k$, in agreement with~\cite{LMRS}.

Explicit calculation of the order $g^2$ radiative corrections to
Eq.~\ref{eq:kk2tree} would not be necessary if, as might be expected
\cite{Park,Osborn:ta}, the mathematical form of the correlation 
function of ${\cal N}=1$ superfields
$\langle S(z_1) \bar{S}(z_2) J(z_3)\rangle$ is uniquely
determined by ${\cal N}=1$ superconformal symmetry as a function of the
three superspace points $z_i = (x_i,\theta_i, \bar{\theta}_i)$, $i= 1,2,3$.
Here $S(z)$ is a chiral operator of dimension $k$, $\bar{S}(z)$ is its
anti-chiral conjugate, and $J(z)$ is a flavor current superfield
containing a conserved vector flavor current $J_{\mu}(x)$ as its
$\theta \bar{\theta}$ component. A unique superconformal form means that
all component correlators, among them
$\langle \tr (z^1)^k \, \tr (\bar{z}^1)^k \, \tr ( \bar{z} t z) \rangle$,
contain a common coupling dependent factor. The same factor occurs
in the 2-point function $\langle \tr(z^1)^k \, \tr (\bar{z}^1)^k\rangle$
because of the Ward identity, and its order $g^2$ term would then be absent
by the calculation of Sec 2. However, current understanding  
of superconformal correlators is still tentative, and we will
therefore present an explicit calculation of the radiative
corrections to $\langle \tr X^k \, \tr X^k \, \tr X^2 \rangle $.

Previously described order $g^2$ interactions that contribute to the
2-point function naturally
appear in the computation of the corrections to the 3-point function.
In addition, there is an interaction that depends on all three space-time
coordinates, which is of the form
\begin{equation}
\label{eq:C}
    \mbox{\raisebox{-.45truecm}{\epsfysize=1.2cm\epsffile{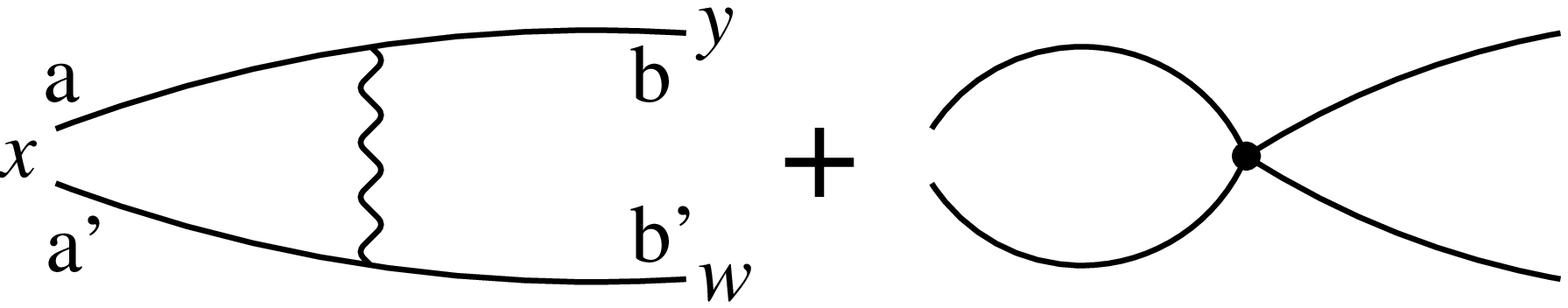} }}
 = (f^{pab} f^{p a' b'} + f^{pab'} f^{p a' b}) C(x;y,w) G(x,y) G(x,w).
\end{equation}
Again, the specific functional form of $C$ will not be needed in the 
argument below.

\begin{figure}[thb]
\centerline{ \epsfysize=3.5cm\epsffile{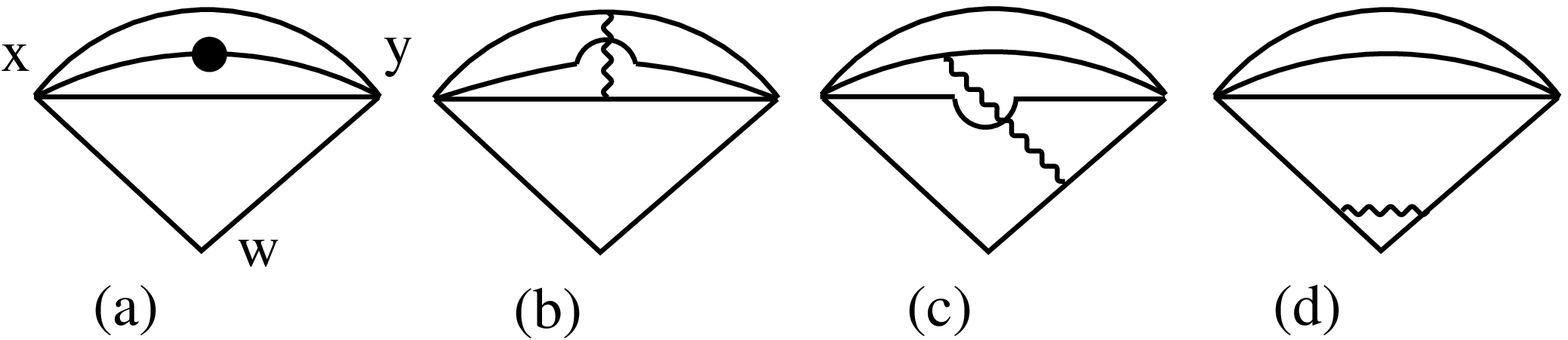}}
\caption{Order $g^2$ corrections to the 3-point function. In addition to gauge
boson exchange diagrams (b), (c), and (d) there are analogous diagrams with
quartic interactions.         \label{fig:kk2}}
\end{figure}
We now analyze combinatorics and color factors present in different
diagrams depicted in Fig.~\ref{fig:kk2}.  The contribution of the
self-energy diagrams is straightforward to evaluate. The sum of
all such diagrams is
\begin{equation}
\label{eq:3A}
 N\left[  (k-1) A(x,y) + A(x,w) + A(y,w)\right]
\end{equation}
where we indicated the space-time arguments of the function $A$ defined
in Eq.~\ref{eq:A}. As usual, we omit the part proportional to the
Born contribution, which is identical for all diagrams. 

Next we consider the insertion of the basic interaction \ref{eq:B}
within the ``rainbow" part of the correlator, as in the diagram
of Fig.~\ref{fig:kk2}(b). We proceed as in Section~2 and insert 
the color factor
into all pairs of lines $i,j$ that belong to the ``rainbow'', and we
replace the $f$ symbols with commutators inside the second trace
$\tr (T^{a_{\sigma(1)}} \cdots T^{a_{\sigma(k)}})$ in
Eq.~\ref{eq:kk2tree}.
We choose the index of the line to the lower vertex to be $a_1$ in the
first trace and
$a_{\sigma(k)}$ in the second, so we sum over $i\neq j=1,...,k-1$.
The sum over the ``rainbow'' lines is
\begin{align}
  - \half  &\sum_{i\neq j=1}^{k-1} \tr ( T^{a_{\sigma(1)}} \cdots
    [T^{a_{\sigma(i)}},T^p] \cdots [T^{a_{\sigma(j)}},T^p]
    \cdots T^{a_{\sigma(k)}})  \nonumber \\
&=  \half \sum_{i=1}^{k-1} \tr(T^{a_{\sigma(1)}} \cdots
   [[T^{a_{\sigma(i)}},T^p],T^p] \cdots T^{a_{\sigma(k)}} ) \nonumber \\
&\qquad {}+\half \sum_{i=1}^{k-1} \tr( T^{a_{\sigma(1)}} \cdots
    [T^{a_{\sigma(i)}},T^p] \cdots [T^{a_{\sigma(k)}},T^p])
\end{align}
where we have used Eq.~\ref{eq:identity}. Using this in the second
term as well as the identity $[[T^a,T^p],T^p]=N T^a$, we can express the above
equation as
\begin{gather}
 \half (k-1) N \, \tr(T^{a_{\sigma(1)}} \cdots T^{a_{\sigma(k)}}) 
  - \half \, \tr(T^{a_{\sigma(1)}} \cdots [[T^{a_{\sigma(k)}},T^p],T^p])
    \nonumber \\
= \half (k-2) N\,  \tr(T^{a_{\sigma(1)}} \cdots T^{a_{\sigma(k)}}).
\end{gather}
The last result has the same color factor as the Born term. After
contracting with the trace for the vertex at $x$, and summing over
permutations, one finds the previous color polynomial $P_{k,k,2}(N)$.
Therefore this set of diagrams yields
\begin{equation}
\label{eq:3B}
  \half (k-2) N B(x,y) P_{k,k,2}(N).
\end{equation}

The remaining order $g^2$ diagrams are obtained by inserting the interaction
of Eq.~\ref{eq:C} into the Born graph. The algebra is quite similar to
the previous case. We first analyze diagrams depicted in Fig.~\ref{fig:kk2}(c).
The color structure of the diagrams with gauge boson exchange
between the single leg and the ``rainbow'' and their analogs with $D$-term
quartic interactions is again obtained by replacing the $f$ symbols
with commutators inside the second trace. One of the commutators
is attached to the distinguished line with the index $a_{\sigma(k)}$, the other
corresponds to a line from the ``rainbow'' $a_{\sigma(i)}$ with
$i=1,\ldots,k-1$:
\begin{align}
 -\half  \sum_{i=1}^{k-1}& \tr (T^{a_{\sigma(1)}} \cdots
    [T^{a_{\sigma(i)}},T^p] \cdots [T^{a_{\sigma(k)}},T^p]) \nonumber \\
&= \half \, \tr (T^{a_{\sigma(1)}} \cdots [[T^{a_{\sigma(k)}},T^p],T^p])
    \nonumber \\
&= \half \, N \tr (T^{a_{\sigma(1)}} \cdots T^{a_{\sigma(k)}}).
\end{align}
Again, the final result is proportional to the Born graph thus the
sum over all permutations is trivial. Both gluon exchange and
quartic interactions from the D-term contribute to the diagrams of
Fig.~\ref{fig:kk2}(c).
Adding the diagrams obtained by symmetrizing in $x,y$, we obtain the result
\begin{equation}
\label{eq:3C}
 2 N [ C(x;y,w) + C(y;x,w)] P_{k,k,2}(N).
\end{equation}

The diagram depicted in Fig.~\ref{fig:kk2}(d) needs to be treated separately,
because quartic interactions which contribute in this case come from both
the D-term and the F-term. We denote the space-time function associated with
the sum of gauge boson and quartic vertices by $C'(w;x,y)$. This is defined
so that $C'$ and $C$ would coincide if F-term vertices are excluded.
All contributions to the diagram have the color factor $N$ from a double
contraction on $ff$ and 
matrix element $t^{11}$ as a flavor factor. The latter is included in the Born 
amplitude (\ref{eq:kk2tree}). It is now straightforward to see that
the net contribution of the graph \ref{fig:kk2}(d) is
\begin{equation}
\label{eq:3Cprime}
  -2 N C'(w;x,y) P_{k,k,2}(N).
\end{equation}

The nonrenormalization theorem for a 3-point function with $k=2$
implies that 
\begin{equation}
\label{eq:3ptnr}
  N \, \bigl[ A(x,y) + A(x,w) + A(y,w) 
+ 2 C(x;y,w) + 2 C(y;x,w) - 2 C'(w;x,y)
  \bigr]=0 .
\end{equation}
We now combine this result with the relation $A+\half B=0$, obtained
from the nonre\-nor\-ma\-liza\-tion of the 2-point function in the 
previous section to show that
there are no order $g^2$ corrections to the 
$\langle \tr X^k \, \tr X^k \, \tr X^2 \rangle $ correlator.
Specifically the sum of all interaction diagrams computed in
(\ref{eq:3A}), (\ref{eq:3B}), (\ref{eq:3C}), and (\ref{eq:3Cprime})
is the Born term (\ref{eq:kk2tree}) multiplied by (\ref{eq:3final})
\begin{align}
\label{eq:3final}
 N &\Bigl[ (k-1) A(x,y) + A(x,w) + A(y,w) + \half (k-2) B(x,y)  
      \nonumber \\ 
&\qquad{}  {} + 2 C(x;y,w) + 2 C(y;x,w) - 2 C'(w;x,y) \Bigr] \nonumber
\\ &= N \Bigl[ (A(x,y) + A(x,w) + A(y,w) + 2 C(x;y,w) 
   + 2 C(y;x,w) - 2 C'(w;x,y))    \nonumber \\
&\qquad{}  + (k-2) ( A(x,y) + \half B(x,y) )  \Bigr]  = 0
\end{align}
and this vanishes by the nonrenormalization theorems for all $k$ and $N$.


          \section{The Correlators \protect\boldmath$\langle \tr X^{k_1} \, \tr
X^{k_2} \,
          \tr X^{k_3} \rangle $\label{sec:ijk}}

In this section we show that a general 3-point function of $\tr
X^k$ does not receive corrections at order $g^2$.  The basic
ingredients of our arguments were outlined in two previous
sections, but the algebra is more complicated for the case at
hand. We first define the symmetric trace as follows
\begin{equation}
  \str (T^{a_1} \cdots T^{a_k}) \equiv \sum_{{\rm perms}\ \sigma}
   \frac{1}{k!} \tr  ( T^{a_{\sigma(1)}} \cdots T^{a_{\sigma(k)}}).
\end{equation}
The Born diagram for the correlator $\langle \tr X^{k_1} \tr
X^{k_2} \tr X^{k_3} \rangle$ contains $\alpha_3 = (k_1 + k_2 -
k_3) / 2$ lines connecting the operators $\tr X^{k_1}$ and $\tr X
^{k_2}$, etc.
It is convenient and sufficiently general to work with highest 
weight components of the first two operators and to choose the
third operator
\begin{displaymath}
  \tr (\bar{z} t z)^{\alpha_1 \alpha_2} 
  = t_{i_1 \cdots i_{\alpha_1 }; j_1 \cdots j_{\alpha_2}}
    \str (\bar{z}^{i_1} \cdots \bar{z}^{i_{\alpha_1}} z^{j_1}
    \cdots z^{j_{\alpha_2}})
\end{displaymath}
where $t$ is any tensor which is symmetric in its $i$ and $j$
indices and traceless upon contraction of an~$i$ and~$j$ index.
This ensures that the operator belongs to an irreducible SU(3)
component of $\tr X^{k_3}$, with $k_3 = \alpha_1 + \alpha_2$.

The free field contribution to the correlator we study is then
\begin{gather*}
\langle \tr (z^1)^{k_1} (x) \tr (\bar{z}^1)^{k_2} (y) \tr
(\bar{z} t z)^{\alpha_1 \alpha_2} (w) \rangle
  = G(x,y)^{\alpha_3} G(y,w)^{\alpha_1}  G(w,x)^{\alpha_2} P_{k_1,k_2,k_3}(N)
  t_{1\cdots1; 1\cdots1}   \nonumber \\
\begin{align}
P_{k_1,k_2,k_3}(N) 
 &= \frac{k_1 ! k_2 !}{ \alpha _3 !}
 \str (T^{a_1}\cdots T^{a_{\alpha_3}} T^{b_1} \cdots T^{b_{\alpha_2}})
    \\
 &\qquad{} \cdot
   \str (T^{b_1}\cdots T^{b_{\alpha_2}} T^{c_1} \cdots T^{c_{\alpha_1}})
   \str (T^{c_1}\cdots T^{c_{\alpha_1}} T^{a_1} \cdots
   T^{a_{\alpha_3}})
   \nonumber
\end{align}
\end{gather*}

There are three classes of diagrams contributing at order $g^2$ to the
correlator. These are self-energy insertions described in Eq.~\ref{eq:A},
gauge boson exchanges within each ``rainbow'' (Eq.~\ref{eq:B}) and gauge boson
exchanges from one ``rainbow'' to another (Eq.~\ref{eq:C}). The self-energy
graphs are straightforward to evaluate. The sum of all such diagrams equals to
\begin{equation}
 N \left[ \alpha_3 A(x,y) + \alpha_2 A(x,w) + \alpha_1 A(y,w) \right]
   P_{k_1,k_2,k_3}(N),
\end{equation}
where we omitted the propagator and flavor matrix factors of the Born 
amplitude.

As the next step we evaluate diagrams with the insertions of the gauge 
boson and $D$-term quartic interactions
described in Eq.~\ref{eq:C} between different ``rainbows.''. 
The color structure of these diagrams is
\begin{gather}
 \sum_{i=1}^{\alpha_3}  \sum_{j=1}^{\alpha_2} (f^{p c_i c'_i}
   f^{p b_j b'_j} + f^{p c_i b'_j} f^{p b_j c'_i}) \Big\{ 
\str (T^{c_1}\cdots T^{c'_i} \cdots
 T^{c_{\alpha_3}} T^{b_1} \cdots T^{b'_j}\cdots T^{b_{\alpha_2}})
\nonumber \\ 
\qquad{}\cdot \str (T^{b_1}\cdots T^{b_{\alpha_2}} T^{a_1} \cdots T^{a_{\alpha_1}})
\,
\str (T^{a_1}\cdots T^{a_{\alpha_1}} T^{c_1} \cdots T^{c_{\alpha_3}})
 \Big\}
\nonumber \\
= -2 \sum_{i=1}^{\alpha_3}  \sum_{j=1}^{\alpha_2}
    \str (T^{c_1} \cdots [T^{c_i},T^p] \cdots  T^{c_{\alpha_3}}
                    T^{b_1} \cdots [T^{b_j},T^p] \cdots  T^{b_{\alpha_2}})
\nonumber \\
\qquad{}\cdot \str (T^{b_1}\cdots T^{b_{\alpha_2}} T^{a_1} \cdots T^{a_{\alpha_1}})
\,
\str (T^{a_1}\cdots T^{a_{\alpha_1}} T^{c_1} \cdots T^{c_{\alpha_3}}).
\end{gather}
We will now show that the color structure arising from insertions of the $f$
symbols on any pairs of indices is identical. 
\begin{align}
 \sum_{i=1}^{\alpha_3}  \sum_{j=1}^{\alpha_2}
    &\str (T^{c_1} \cdots [T^{c_i},T^p] \cdots  T^{c_{\alpha_3}}
                    T^{b_1} \cdots [T^{b_j},T^p] \cdots  T^{b_{\alpha_2}})
\nonumber \\ &\qquad{}\cdot 
\str (T^{b_1}\cdots T^{b_{\alpha_2}} T^{a_1} \cdots T^{a_{\alpha_1}})
\,
\str (T^{a_1}\cdots T^{a_{\alpha_1}} T^{c_1} \cdots T^{c_{\alpha_3}})
\nonumber  \\ &=
\str (T^{c_1} \cdots T^{c_{\alpha_3}} T^{b_1} \cdots   T^{b_{\alpha_2}})
\sum_{j=1}^{\alpha_2} \str (T^{b_1}\cdots [T^{b_j},T^p] \cdots  T^{b_{\alpha_2}}
                 T^{a_1} \cdots T^{a_{\alpha_1}})
\nonumber \\ &\qquad{}\cdot 
\sum_{i=1}^{\alpha_3} \str (T^{c_1}\cdots T^{c_{\alpha_1}}
                      T^{c_1} \cdots [T^{c_i},T^p] \cdots  T^{c_{\alpha_3}})
\nonumber \\ &=
-\str (T^{c_1} \cdots T^{c_{\alpha_3}} T^{b_1} \cdots \cdots  T^{b_{\alpha_2}})
\sum_{k=1}^{\alpha_1} \str (T^{b_1}\cdots T^{b_{\alpha_2}}
                      T^{a_1} \cdots [T^{a_k},T^p] \cdots  T^{a_{\alpha_1}})
\nonumber \\ &\qquad{}\cdot 
\sum_{i=1}^{\alpha_3} \str (T^{a_1}\cdots T^{a_{\alpha_1}}
                      T^{c_1} \cdots [T^{c_i},T^p] \cdots  T^{c_{\alpha_3}})
\nonumber \\ &=   
\str (T^{c_1} \cdots T^{c_{\alpha_3}} T^{b_1} \cdots \cdots  T^{b_{\alpha_2}}) 
\, 
\str (T^{b_1}\cdots T^{b_{\alpha_2}} T^{a_1} \cdots T^{a_{\alpha_1}})
\nonumber \\ &\qquad{}\cdot 
\sum_{i=1}^{\alpha_3} \sum_{k=1}^{\alpha_1} 
\str (T^{a_1} \cdots [T^{a_k},T^p] \cdots  T^{a_{\alpha_1}}
                T^{c_1} \cdots [T^{c_i},T^p] \cdots  T^{c_{\alpha_3}}).
\end{align}
Using this last result we may now define the above quantity as a function
$W_{k_1,k_2,k_3} (N)$, which is independent of the location of insertion of
the two commutators. Adding all three sets of diagrams with interactions
among two different ``rainbows'', we obtain
\begin{equation}
 -2 W_{k_1,k_2,k_3}(N) \left[ C(x;y,w) + C(y;x,w) - C(w;x,y) \right] ,
\end{equation}
since the color factors are identical for the contributions associated with
each of the three vertices, except for a sign at the $\tr X^{k_3}$ vertex
because the operator contains both $z$ and $\bar{z}$ fields.

The next step is the evaluation of diagrams with gauge boson exchanges
within one ``rainbow''. The color structure of such diagrams is of the
form
\begin{align}
   -\half \sum_{i\neq j=1}^{\alpha_2} &\str (T^{c_1} \cdots T^{c_{\alpha_3}}
              T^{b_1}\cdots [T^{b_i},T^p] \cdots [T^{b_j},T^p] \cdots 
T^{b_{\alpha_2}})
       \, \str (\cdots )  \, \str (\cdots ) \nonumber \\
 &=  \half \Bigl\{ \sum_{i=1}^{\alpha_3} \sum_{j=1}^{\alpha_2}
              \str (T^{a_1} \cdots [T^{a_i},T^p] \cdots  T^{a_{\alpha_3}} 
                    T^{b_1} \cdots [T^{b_j},T^p] \cdots  T^{b_{\alpha_2}})
\nonumber \\
 &\qquad{}  + \sum_{j=1}^{\alpha_2} \str (T^{a_1} \cdots T^{a_{\alpha_3}}
                    T^{b_1} \cdots [[T^{b_j},T^p],T^p] \cdots  T^{b_{\alpha_2}})
        \Bigr\} \, \str (\cdots )  \, \str (\cdots ) \nonumber \\
 &= \half W_{k_1,k_2,k_3}(N) + \half N \alpha_2 P_{k_1,k_2,k_3}(N).   
\end{align}

Finally, we evaluate the contribution of the quartic $F$-term vertex in the
Lagrangian 
~\ref{eq:n4lag}. There are nonvanishing Wick-contractions with one $z$ and
one $\bar{z}$ field of the operator $\tr X^{k_3}$, 
and one line is attached to each of the two other
vertices. One can see by contracting the $\epsilon$ symbols in the
F-term of ~\ref{eq:n4lag}, using the tracelessness of $t$, that this 
contribution also has the flavor factor
$t _{1\cdots 1 ; 1 \cdots  1}$,
The space-time function associated with the F-term will combine with
gauge boson and D-term contributions at the $\tr X^{k_3}$ vertex to
give the same function $C'(w;x,y)$ used in our discussion of the
$k-k-2$ correlator in section 3. 
It remains to determine the
color structure of this contribution, which is again obtained from the 
quartic 
Lagrangian vertex. The result is
\begin{gather}
\half \sum _{i=1}^{\alpha _1} \sum _{j=1} ^{\alpha _2}  
\str (T^{a_1} \cdots T^{a_i} \cdots T^{a_{\alpha _1}}
     T^{b_1} \cdots T^{b_j} \cdots T^{b_{\alpha _2}})
f^{p a_i' b_j} f^{p b_j ' a_i}
\nonumber \\
  \str (T^{c_1} \cdots T^{c_{\alpha _3}} T^{b_1} \cdots
   T^{b'_j} \cdots T^{\alpha _2} )
\str (T^{c_1} \cdots T^{\alpha _3} T^{a_1} \cdots T^{a'_i} \cdots
T^{\alpha _1} ) 
\end{gather}
Replacing $f^{p a_i ' b_j} T ^{b_j} = -i [ T^p, T^{a_i'}]$ and
$f^{p b_j' a_i} T^{a_i} = -i [ T^p,  T^{b_j '}]$, we 
recognize this double sum as proportional to the quantity
$W_{k_1,k_2,k_3}(N)$, introduced earlier.

We can now combine all contributions together to obtain
\begin{gather} 
N \Bigl[ \alpha_3 (A+\half B)(x,y) + \alpha_2 (A+\half B)(x,w)
     + \alpha_1 (A+ \half B)(y,w)
           \Bigr] P_{k_1,k_2,k_3}(N) \nonumber  \\
 - \Bigl[2\Bigl( C(x;y,w) + C(y;x,w) - C'(w;x,y)\Bigr) - \half
\Bigl( B(x,y) -  B(x,w) - B(y,w)\Bigr) \Bigr]
               W_{k_1,k_2,k_3}(N).
\end{gather}
This can be seen to vanish when the nonrenormalization theorems used in 
previous sections are applied, specifically when $A +\half B=0$ is combined
with  Eq.~\ref{eq:3final}.
This result shows that order $g^2$ corrections
cancel for arbitrary $N$ in all 3-point functions of the operators $\tr X^k$.

\section{A descendent correlator\label{sec:descendent}}
\setcounter{equation}{0}

It is commonly believed that the 3-point correlators of chiral 
primary operators determine those of descendents. For ${\cal N}=1$
SUSY the form of superconformal correlators has been discussed by
\cite{1001,Park,Osborn:ta}. If there
is a unique super-conformal ``tensor'' form for a given superfield
correlator, then the correlation functions of all components are
determined. For ${\cal N}=4$ SUSY no off-shell superspace (or auxiliary field)
formalism is known, but there is an on-shell formalism \cite{HoweWest}
which predicts a unique 3-point supercorrelator for the multiplet
containing the primary $\tr X^2$ and descendents. Other useful results
have been deduced within this formalism \cite{Ferrara}, but the rules of
applicability are not completely clear (at least to us), so it is
desirable to check its predictions. Absence of radiative corrections
is known for the 3-point function of flavor currents~\cite{MIT}
and for the stress tensor~\cite{Osborn:pc}, and we shall study
here an amplitude involving the second descendent of $\tr X^2$.

We shall discuss this descendent from an ${\cal N}=1$ viewpoint. 
For a chiral multiplet $(z,L \psi,F)$, where $\psi$ is a Majorana spinor and
$L$ is the left chiral projector, the second descendent of $z^2$ is the
operator $S = \bar{\psi} L\psi - 2zF$. For the Lagrangian (\ref{eq:n4lag})
the ${\cal N}=1$ auxiliary fields are $F_i^a = \half Y f^{abc} \epsilon_{ijk}
z^j_b z^k_c$. This defines the descendent operator
\begin{equation}
\label{eq:Sij}
  S^{ij}=\bar{\psi}^i_a L \psi^{j}_a -\half Y f^{abc}
         \epsilon^{mn \{i } z_a^{j\} } \bar{z}^m_b \bar{z}^n_c
\end{equation} 
which transforms in the ${\bf 6}$ representation of $SU(3)$ flavor,
which is contained in the decomposition of the ${\bf 10}$ of $SU(4)$,
whose components can be denoted by $S^{ij}, i,j = 1,\ldots,4$.
We will study the correlator $\langle S^{ij} S^{kl} \tr X^2 \rangle$
which vanishes in free field approximation but could
have an order $g^2$ contribution. Indeed, there is one $SU(4)$ invariant
coupling of ${\bf 10} \otimes {\bf 10} \otimes {\bf 20'}$ representations, so to
check for the absence of radiative corrections, it is sufficient to choose the
convenient  set of components $\langle S^{44}(x) S^{11}(y) \tr (\bar{z}^1(w)) ^2
\rangle$.
Here, $S^{44}$ is the $SU(4)$ partner of $S^{ij}$ whose bi-fermion term
is $\tr \bar{\lambda} L \lambda$, where $\lambda$ is the gaugino of the
${\cal N}=1$ description. To find the correctly
normalized tri-boson term we use the fact that the ${\bf 10}$ representation
is both the symmetric second rank tensor in the $SU(4)$ description and
the self-dual third rank tensor of the $SO(6)$ description of flavor.
We use an $SO(6)$ Clifford algebra construction\footnote{
In six Euclidean dimensions there are 6 Hermitean $8\times 8$
gamma matrices $\Gamma^i$ and a chirality matrix $\bar \Gamma =i \Gamma ^1 
\Gamma ^2 \Gamma ^3 \Gamma ^4 \Gamma ^5 \Gamma ^6$.
There is a symmetric charge conjugation
matrix $C$, which anti-commutes with $\bar \Gamma$. Let $\Gamma^{ijk}$ denote the
antisymmetric third rank tensor of the Clifford algebra. The $\Gamma^{ijk}C$ are
symmetric  matrices, while $(1 \pm \bar \Gamma) \Gamma^{ijk}C$ are symmetric,
self-dual, and effectively $4\times 4$. The contraction of
these matrices with $\tr X^i X^j X^k$ produces a symmetric second rank $SU(4)$
tensor. The relative normalization of the bosonic admixtures in  
Eqs.~(\ref{eq:Sij}) and (\ref{eq:S44}) may be obtained from
a specific construction of these matrices.}
to connect these descriptions.
Omitting tedious details, the result for the descendent is
\begin{equation}
\label{eq:S44}
  S^{44}=\bar{\lambda}_a L \lambda_a - \third Y f^{abc} \epsilon_{ijk}
  z^i_a z^j_b z^k_c .
\end{equation}
(Of course, to enforce ${\cal N}=4$ SUSY, we should set $Y=\sqrt{2} g$, but 
 we postpone this until the end of the calculation.)

\begin{figure}[htb]
\centerline{\epsfxsize=\textwidth \epsffile{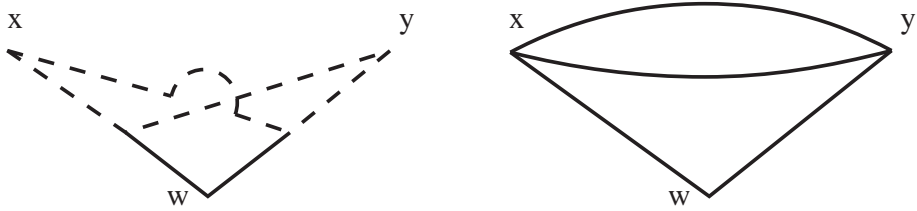}}
  \vspace{-6.8in}
\caption{Two graphs contributing to the
  $\langle S^{44} S^{11} \tr (\bar{z}^1)^2 \rangle$ correlator.
  Solid lines represent the boson propagators, broken ones fermion
  propagators.
         \label{fig:SSz}}
\end{figure}
There are two diagrams for the 2-loop correlator 
$\langle S^{44}(x) S^{11}(y) \tr (\bar{z}^1(w))^2 \rangle$,
a nonplanar diagram with a fermion loop and a purely bosonic Born-like
diagram. Diagrams are drawn in Fig.~\ref{fig:SSz}.
The 8-dimensional integral in the nonplanar amplitude
can be evaluated using the technique of conformal
inversion~\cite{BakerJohnson,JoshDan}, and the other amplitude is elementary.
Attention to combinatoric factors is essential. The results are
\begin{align}
 {\rm nonplanar} &= -g^2 N (N^2-1) G(x,y)^2 G(x,w) G(y,w)
                      \\
 {\rm bosonic}    &= \half Y^2 N (N^2-1) G(x,y)^2 G(x,w) G(y,w) .
\end{align}
We see that these two graphs cancel one another only when the ${\cal N}=4$
relation $Y=\sqrt{2}g$ is imposed, so this is clearly not a consequence of just
${\cal N}=1 $ SUSY.

One should note that each graph is itself the lowest order
contribution to a 3-point correlator of gauge invariant operators
in the ${\cal N}=4$ SYM theory. In each amplitude only two of the three 
operators
have scale dimension protected by R-symmetry. These correlators are
thus not accessible from $AdS_5$ calculations with supergravity or 
Kaluza-Klein fields. Perhaps they can be computed using appropriate
string modes and interactions.

\section{Conclusions}
The main result of this paper is that order $g^2$ radiative
corrections to the correlators $\langle \tr X^k\, \tr X^k \rangle$ and 
$\langle \tr X^{k_1} \,  \tr X^{k_2} \, \tr X^{k_3} \rangle$
cancel to all orders in $N$.  This confirms the $N \rightarrow \infty$
fixed large $g^2N$, AdS$_5$ supergravity calculation of Ref.~\cite{LMRS}
and the additional conjecture and speculation made in that paper concerning
finite $N$ behavior. The result goes beyond known nonrenormalization
theorem for 2- and 3-point functions of $\tr X^2$ and thus suggests an
unsuspected additional simplicity or symmetry of the ${\cal N}=4$ SYM theory.

Our method used careful combinatorics to reduce the interaction terms
of general $\tr X^k$ correlators to multiples of those of $\tr X^2$, which
vanish by nonrenormalization theorems. It seems unlikely that the method 
can be applied
in higher order. For example in order $g^4$ there are interactions which
involve the exchange of two gluons among three adjacent scalar lines of the
``rainbow'' of Fig.~\ref{fig:2Born}.
It does seem clear, though, that the method and the order $g^2$ result are valid
for the ${\cal N}=4$  theory with any gauge group.

The Montonen-Olive duality of ${\cal N}=4$ SYM theory relates weak and strong
coupling behavior of the theory at fixed $N$. The
gauge invariant correlators we study are invariant under the S-duality
group so the absence of order $g^2$ radiative corrections for small $g$
implies  that there are no order $1/g^2$ terms in the strong coupling 
expansion. This is valid for all $N$ and therefore consistent with but
stronger than the conclusions of Ref.~\cite{BanksGreen}. In this paper it was 
argued that the lowest order nonperturbative string corrections to the
type IIB effective supergravity action are of order $\alpha^{\prime3} R^4$ where
$\alpha'$ is the string tension and $R^4$ denotes a quartic contraction of 
the 10-dimensional curvature
tensor. This correction term therefore makes no contribution to 2- and
3-point boundary correlators of the graviton and its supersymmetric partners,
including Kaluza-Klein modes, in the $AdS_5\times S_5$ theory \cite{vanN}.
Using the Maldacena correspondence, a cubic correction of the form 
$\alpha^{\prime2} R^3$ would produce order $1/g^2N$ corrections as
$N\rightarrow \infty$ in the correlators
$\langle \tr X^{k_1} \,  \tr X^{k_2} \, \tr X^{k_3} \rangle$.

\setcounter{equation}{0}

\section*{Acknowledgments}
We thank Massimo Bianchi, Marc Grisaru, Paul Howe, Shiraz Minwalla, Leonardo
Rastelli, and Daniela Zanon for useful discussions during the course of this work.
The research of E.D'H. is supported in part by National Science Foundation under
Grant No.\ PHY-95-31023, D.Z.F is supported in part by NSF Grant No.\
PHY-97-22072. W.S. is supported in part by Department of Energy  under Grant
No.\ DOE-FG03-97ER40506.

\end{document}